\documentclass[pra,aps,twocolumn,floatfix,showpacs,tightenlines,superscriptaddress,amsmath,amssymb,footinbib]{revtex4}
\usepackage{amssymb,amsbsy,epsfig,color,graphicx,bbm,amsmath}

\usepackage[latin1]{inputenc}
\newcommand{\sig}{\gr\sigma}
\newcommand{\id}{\mathbbm{1}}
\newcommand{\R}{\mathbbm R}
\newcommand{\gr}[1]{\boldsymbol{#1}}
\newcommand{\be}{\begin{equation}}
\newcommand{\ee}{\end{equation}}
\newcommand{\bea}{\begin{eqnarray}}
\newcommand{\eea}{\end{eqnarray}}

\renewcommand{\det}{{\rm Det}\,}
\newcommand{\eq}[1]{Eq.~(\ref{#1})}
\newcommand{\tr}[1]{\,{\rm Tr}\,#1}

\begin{document}

\title{Geometric characterization of separability and entanglement in pure \\
       Gaussian states by single-mode unitary operations}

\date{September 4, 2007}

\author{Gerardo Adesso}
\affiliation{Dipartimento di Fisica ``E. R. Caianiello'',
Universit\`a degli Studi di Salerno, Via S. Allende, I-84081
Baronissi (SA), Italy}
\affiliation{CNR-INFM Coherentia, Napoli, Italy; CNISM, Unit\`a di Salerno; and INFN,
Sezione di Napoli - Gruppo Collegato di Salerno, Italy}
\affiliation{Grup d'Informació Quàntica, Universitat Autònoma de Barcelona, E-08193
Bellaterra (Barcelona), Spain}

\author{Salvatore M. Giampaolo}
\affiliation{Dipartimento di Matematica e
Informatica, Universit\`a degli Studi di Salerno, Via Ponte don
Melillo, I-84084 Fisciano (SA), Italy}
\affiliation{CNR-INFM Coherentia, Napoli, Italy; CNISM, Unit\`a di Salerno; and INFN,
Sezione di Napoli - Gruppo Collegato di Salerno, Italy}

\author{Fabrizio Illuminati}
\thanks{Corresponding author. Electronic address: illuminati@sa.infn.it}
\affiliation{Dipartimento di Matematica e
Informatica, Universit\`a degli Studi di Salerno, Via Ponte don
Melillo, I-84084 Fisciano (SA), Italy}
\affiliation{CNR-INFM Coherentia, Napoli, Italy; CNISM, Unit\`a di Salerno; and INFN,
Sezione di Napoli - Gruppo Collegato di Salerno, Italy}
\affiliation{ISI Foundation for Scientific Interchange, Viale Settimio Severo
65, I-10133 Turin, Italy}

\pacs{03.65.Ud, 03.67.Mn}

\begin{abstract}
We present a geometric approach to the characterization of separability
and entanglement in pure Gaussian states of an arbitrary number of
modes. The analysis is performed adapting to continuous variables a
formalism based on single subsystem unitary transformations that has
been recently introduced to characterize separability and entanglement
in pure states of qubits and qutrits [arXiv:0706.1561]. In analogy with
the finite-dimensional case, we demonstrate that the $1 \times M$
bipartite entanglement of a multimode pure Gaussian state can be quantified
by the minimum squared Euclidean distance between the state itself and
the set of states obtained by transforming it via suitable local
symplectic (unitary) operations. This minimum distance, corresponding
to a, uniquely determined, extremal local operation, defines a novel
entanglement monotone equivalent to the entropy of entanglement, and
amenable to direct experimental measurement with linear optical schemes.
\end{abstract}

\maketitle

\section{Introduction}

The concept of entanglement has gradually developed from the status
of a puzzling interpretational problem, to that of a crucial operational
resource for quantum information tasks and, even more remarkably, to the
status of a founding property of quantum theory, whose implications and
applications extend into many diverse areas of research ranging from
quantum optics and atomic and molecular physics to condensed matter
physics and quantum critical phenomena \cite{reviewfazio,reviewhoro}.
While many open questions, even on defining grounds, stand open when
it comes to address questions like the nature of multipartite entanglement
and the entanglement of mixed states, a fairly satisfactory classification
and quantification of bipartite entanglement of pure quantum states have
been established \cite{reviewhoro,pleniovirmani}. This achievement has been
partly possible because the milestones of quantum information
science, like quantum teleportation, quantum cryptography, state transfer,
broadcasting and telecloning, entanglement creation and distillation, all
rely on the paradigm of two distant labs operated by two parties -- traditionally
named Alice and Bob -- who wish to communicate, possibly sharing a pure
entangled state \cite{nielsenchuang}. By properly defining figures
of merit associated with such protocols, pure-state bipartite
entanglement has been understood both qualitatively -- entangled states
are non-separable -- and quantitatively -- the degree of pure-state
bipartite entanglement is equal to the entropy of the reduced state
of each subsystem. In particular, the von Neumann entropy of
entanglement is equal both to the, operationally defined, distillable
entanglement and entanglement cost of pure bipartite quantum states \cite{bennett}.
This equivalence is lost in the presence of mixedness, and the
phenomenon of entanglement conversion irreversibility
sets in \cite{irrevers}.

There exists an alternative, equally natural way to understand
and characterize entanglement. From a {\em geometric} perspective,
the degree of entanglement in a state $\rho$ can be quantified as
the minimum distance, suitably measured, between $\rho$ and the set
of unentangled, separable states \cite{pleniovirmani}. Again, for
pure states, if such distance is measured in terms of the relative
entropy, the resulting measure of entanglement coincides with the
von Neumann entropy of entanglement \cite{vedral}. This suggests
that other entanglement monotones, that can be useful either for their
operational meaning and/or for their computability, might be endowed
with an alternative, geometric interpretation which adds to their
understanding and may provide alternative tools in their
experimental detection. A novel approach to the study of
this problem has been recently introduced for low-dimensional
discrete-variable systems such as qubits and qutrits \cite{squo}.
It relies on the basic idea that entanglement can be characterized
by the response of a system to {\em local} and {\em unitary}
perturbations that, by definition, cannot change the degree of
entanglement present in the system. Notwithstanding this simple fact,
oddly enough, degrees of freedoms that are affected by local unitaries
tend to be systematically neglected in the analyses of entanglement
properties \cite{linden-gerry}. The recent study by Giampaolo and Illuminati
\cite{squo} yields instead that there exist specific single-party unitary
operations (corresponding to the projection on the $\hat z$ component of
the spin for qubits and qutrits) which have the following properties:
(i) they leave a pure bipartite state invariant if and only if it is a product
state; (ii) they transform any pure bipartite entangled state in such a way that
the minimum squared Euclidean distance of the original state from the
set of all possible transformed states is an entanglement monotone.
In the case of bipartite states of qubits and qutrits, such a measure coincides
exactly with the marginal linear entropy, quantifying the degree
of impurity (mixedness) of each subsystem \cite{squo}. Therefore, entanglement
monotones based on completely different definitions, such as the
linear entropy and the tangle \cite{ckw}, are re-discovered and re-interpreted
in terms of the Hilbert-space distance between quantum states and their images
under suitably selected local unitary operations.

In this work, we apply the framework introduced in
Ref.~\cite{squo} to characterize entanglement of pure Gaussian
states of continuous-variable systems. Recent progresses have
showed that many nontrivial problems in entanglement theory,
whose remarkable complexity renders their solution unachievable in
qu$d$it systems with $d$ greater than $2$ or $3$, can be
successfully tackled with different techniques when considering
systems defined on infinite-dimensional Hilbert spaces, like, e.g.
the quantum electromagnetic field \cite{reviewcv}. In particular,
Gaussian states, such as coherent, squeezed states, and in general
all ground and thermal states of harmonic lattices, have played an
increasingly important role in quantum information science,
thanks to their simple structural properties as well as to the high
degree of experimental control on their production and manipulation
\cite{brareview,librocvcerf}. Motivated by these considerations, we
seek here to provide a novel geometric interpretation for bipartite
entanglement of pure Gaussian states, in terms of the perturbation
induced on them by single-mode unitary operations in Hilbert space,
or, equivalently, symplectic transformations in
quantum phase space. We will find, in direct
analogy with the discrete-variable case \cite{squo}, that there
exists a single-mode symplectic operation which preserves product
states, and whose action leads in general to the definition of a
pure-state entanglement monotone for $1 \times N$ Gaussian states.
This measure does not exactly coincide with any known entanglement
measure, even though it is a monotonically increasing function of
the entropy of entanglement, providing thus a novel quantifier of
continuous variable entanglement endowed with a purely geometric
interpretation.

The paper is organized as follows: in Section II we briefly review
the basic tools of the symplectic formalism in phase space, that is
best suited for the analysis of separability, entanglement, and
quantum operations on Gaussian states of infinite-dimensional
quantum systems. In section III we introduce and analyze the
properties of single-mode unitary (symplectic) operations in quantum
phase space, and define the distance, induced by the fidelity,
between pure Gaussian states and their images under
such operations (these images are again pure Gaussian states). We
then proceed to determine the minimum distance over the set of all
possible such transformations, and the associated {\em extremal}
operation. We prove that invariance of a state under the action of
the extremal operation is a necessary and sufficient condition for
the full separability of multimode pure Gaussian states of
translationally invariant system, and show that the associated
minimum distance is an entanglement monotone closely related to the
linear entropy of the subsystem reductions. We finally discuss the
relation between this novel entanglement monotone and the various
possible extensions of the definition of the tangle to continuous
variable systems. In Section IV we point out at some possible future
lines of investigation in the framework of the formalism of local
symplectic operations, also concerning mixed states, and discuss
possible methods for the direct experimental detection of the
minimum distance using linear optical elements, with an explicit
example focused on tripartite Gaussian states.

\section{Phase-space description of Gaussian states and single-mode symplectic operations}

We consider a continuous-variable (CV) system consisting of $N$
canonical bosonic modes, associated with an infinite-dimensional
Hilbert space, tensor product of the $N$ single-mode Fock spaces
\cite{reviewcv,brareview,reviewjens}. Unitary operations which are
at most quadratic in the canonical operators, amount to symplectic
transformations in phase space. A real $2N \times 2N$ matrix
describes a  symplectic transformation  $S \in Sp(2N,\R)$ if, by
definition, it preserves the symplectic form,
\begin{equation}
\label{sy}
S \Omega S^T = \Omega,\quad \Omega=\omega^{\oplus N},\quad
\omega=\left(
                                        \begin{array}{cc}
                                          0 & 1 \\
                                          -1 & 0 \\
                                        \end{array}
                                      \right)\,.
\end{equation}
For a single mode, the generators of the symplectic group $Sp(2,\R)$
are \cite{arvind}
\begin{equation}\label{gen}
\Sigma_1 = \left(
                                        \begin{array}{cc}
                                          0 & 1 \\
                                          1 & 0 \\
                                        \end{array}
                                      \right),\,\,
\Sigma_2 = \left(
                                        \begin{array}{cc}
                                          0 & 1 \\
                                          -1 & 0 \\
                                        \end{array}
                                      \right),\,\,
\Sigma_3 = \left(
                                        \begin{array}{cc}
                                          1 & 0 \\
                                          0 & -1 \\
                                        \end{array}
                                      \right),
\end{equation}
where $\Sigma_2 = \omega$. The matrices $\Sigma_i$'s
in \eq{gen} are traceless. Together with the
identity matrix $\id$, they form a basis in the space of $2\times2$
real matrices. According to the Euler decomposition, the most
general single-mode symplectic operation $S \in Sp(2,\R)$ can be
written as a sequence of a rotation, a  squeezing, and a second
rotation (with different angle) in phase space,
\begin{equation}\label{syrsr}
S=\left(
\begin{array}{cc}
 \cos \phi & \sin \phi \\
 - \sin \phi & \cos \phi
\end{array}
\right)\! \left(\begin{array}{cc}
 \xi &  0 \\
 0 & {\xi}^{-1}
\end{array}
\right)\! \left(\begin{array}{cc}
 \cos \theta & \sin \theta  \\
 - \sin \theta & \cos \theta
\end{array}
\right)\!,
\end{equation}
reducing to the identity transformation for $\theta=\phi=0,\,\xi=1$.

We are interested in studying the minimal distance between a state
and its image as transformed by a specific type of local single-mode
symplectic operations. Clearly, one cannot allow the identity
transformation in the defining set of possible operations, if one
wants to avoid ending up with a trivial null distance on all
quantum states. Then, in analogy with the finite-dimensional case,
we impose the condition
of tracelessness \cite{squo}, and we define a {\em unitary single-mode operation}
$S_{smo}$ as the most general $Sp(2,\R)$ symplectic transformation
of the form \eq{syrsr}, with $\tr{S}=0$. In this way we are only
considering symplectic transformations which are orthogonal to the
identity. Imposing such constraint yields $\phi=\pi/2-\theta$,
namely
\begin{equation}\label{suno}
S_{smo} = \left(
\begin{array}{cc}
\frac{\left(\xi^2 - 1\right) \cos \theta  \sin \theta }{\xi} & \frac{\cos ^2 \theta + \xi^2 \sin ^2 \theta}{\xi} \\
- \frac{\xi^2 \cos ^2 \theta + \sin ^2 \theta}{\xi} & - \frac{\left(\xi^2 -
1\right) \cos \theta \sin \theta }{\xi}
\end{array}
\right) .
\end{equation}
The transformation $S_{smo}(\xi,\theta)$ can be written as a linear combination
of the $\Sigma_i$'s from \eq{gen}, $S_{smo} = \alpha \Sigma_1 +
\beta \Sigma_2 + \gamma \Sigma_3$, where the symplectic condition
\eq{sy} imposes $\gamma=\sqrt{\beta^2 -\alpha^2 - 1}$, $\beta \ge
\sqrt{\alpha^2+1}$. Explicitly \footnote{\eq{m} can be also seen as
an infinitesimal symplectic transformation, obtained from the
first-order expansion of $S=\exp[\alpha \Sigma_1 + \beta \Sigma_2 +
\gamma \Sigma_3] \simeq \id + (\alpha \Sigma_1 + \beta \Sigma_2 +
\gamma \Sigma_3)$, minus the identity $\id$.}:
\begin{equation}
\label{m}
S_{smo}(\alpha,\beta)=\left(
\begin{array}{cc}
 \sqrt{ \beta ^2 -\alpha^2 - 1} & \alpha + \beta  \\
 \alpha - \beta  & - \sqrt{ \beta ^2-\alpha ^2 - 1}
\end{array}
\right)\,,
\end{equation}
where the parameters $\alpha,\beta$ are connected with the squeezing
$\xi$ and the rotation angle $\theta$, appearing in \eq{suno}, by
the following relations:
\begin{eqnarray}
\label{stheta}
\xi &=& [\beta  \alpha + \alpha \sqrt{\beta ^2 - 1}]/{\alpha }\,, \nonumber \\
&& \\
\cos \theta &=& \sqrt{[\beta ^2 - \alpha  \sqrt{\beta ^2 - 1} -
1]/[2(\beta ^2 - 1)]}\,. \nonumber
\end{eqnarray}

\section{Extremal single-mode operations and entanglement of pure Gaussian states}

We can now move to the specific setting of the geometric analysis.
Let our $N$-mode bosonic system be prepared in a pure Gaussian state
\cite{reviewcv}. We recall that Gaussian states of $N$ modes are
completely described in phase space (once the first moments are set
to zero via local displacements) by the real, symmetric $2N\times
2N$ covariance matrix (CM) $\gr{\sigma}$, whose entries are
$\sigma_{lm}=1/2\langle\{\hat{X}_l,\hat{X}_m\}\rangle
-\langle\hat{X}_l\rangle\langle\hat{X}_m\rangle$. Here
$\hat{X}=\{\hat x_1,\hat p_1,\ldots,\hat x_N,\hat p_N\}$ is the
vector of the field quadrature operators, whose canonical
commutation relations can be expressed in matrix form: $[\hat
X_{l},\hat X_m]=2i\,\Omega_{lm}$, with the symplectic form $\Omega$
defined in \eq{sy}. According to Williamson theorem
\cite{williamson36}, the CM of a $N$-mode Gaussian state can be
always diagonalized by means of a global symplectic transformation
(this corresponds to the normal mode decomposition): $W_{\sig} \sig
W_{\sig}^T = \gr\nu$, where $W_{\sig}\in Sp(2N,\R)$ and
$\gr{\nu}=\bigoplus_{k=1}^{N}{\rm diag}\{\nu_k,\,\nu_k\}$ is the CM
corresponding to the tensor product of single-mode thermal states.
The quantities $\{\nu_k\}$ are the so-called symplectic eigenvalues
of the CM $\sig$.

A {\em pure} Gaussian state is characterized by $\nu_k=1$,
$\forall\,\,k=1\ldots N$, which implies $\det\sig=1$. Such a state
may be, for instance, the ground state of some harmonic Hamiltonian.
We want to study the $1 \times (N-1)$ entanglement of one mode with
the remaining $N-1$ modes, via the perturbation induced by
single-mode operations on mode $1$. Namely, we aim to study the
minimal squared distance between the Gaussian state $\sig$ and the
state obtained from it by applying a $S_{smo}$ of the form \eq{m} on
any selected mode, for instance mode $1$. It is important to recall
that the transformed state, being obtained from the original pure
Gaussian state by applying to it a symplectic transformation, i.e.~a
unitary transformation at most quadratic in the field variables, is
again a pure Gaussian state. Introducing the standard Bures metric,
the minimum distance reads
\begin{equation}
\label{distanza}
D(\sig)=\min_{\alpha,\beta}\left[1-{\cal F}(\sig,\sig')\right]\,.
\end{equation}
Here $\sig'=[S_{smo}(\alpha,\beta)\oplus \id_{2\ldots N}]\cdot \sig
\cdot [S_{smo}(\alpha,\beta)\oplus \id_{2\ldots N}]^T$, and the {\em
fidelity} ${\cal F}$ between two pure-state $N$-mode Gaussian CMs
can be computed as \cite{marian} $${\cal F}(\sig,\sig')
=2^N/\sqrt{\det{(\sig+\sig')}},$$ amounting to the overlap
$|\langle\psi|\psi'\rangle|^2$ between the original and the
perturbed Gaussian quantum states.

To proceed in the evaluation of \eq{distanza}, let us first
take into account that, in full generality,
pure Gaussian states can always be brought in the phase-space Schmidt form
\cite{reviewcv} with respect to the $1 \times (N-1)$ bipartition. The
symplectic transformation $W$ achieving the Schmidt decomposition is
the direct sum of the two Williamson diagonalizing operations acting
on the single-mode and the $(N-1)$-mode subspaces, respectively, $W
= W_1 \oplus W_{2\ldots N}$. One then has
\begin{eqnarray}
\label{schm}
\sig_W &=& W \sig W^T  \nonumber \\
&& \nonumber \\
&=& \left(
\begin{array}{cccc}
a & 0 & \sqrt{a^2 - 1} & 0 \\
0 & a & 0 & - \sqrt{a^2 - 1} \\
\sqrt{a^2 - 1} & 0 & a & 0 \\
0 & - \sqrt{a^2 - 1} & 0 & a
\end{array}
\right) \nonumber \\
&& \nonumber \\
& & \oplus \, \, \id_{3\ldots N} \, , \quad \mbox{with
$a\ge1$}\, ,
\end{eqnarray}
i.e.~the phase-space Schmidt form of $\sig$ is constituted by one
two-mode squeezed state between modes 1 and 2, tensor $N-2$
uncorrelated vacua \cite{boteroecc}. To evaluate \eq{distanza}, we
need the minimum of $\det[\sig+(S_{smo}\oplus \id_{2\ldots N})\sig
(S_{smo}\oplus \id_{2\ldots N})]^T$. We will now show that it is
enough to consider states in the form $\sig_W$. In fact,
\begin{widetext}
\begin{eqnarray}
\label{longproof}
&& \det[\sig_W+(S_{smo}\oplus \id_{2\ldots N})\sig_W
(S_{smo}\oplus \id_{2\ldots N})^T]  \nonumber \\
&=& \det[(W_1 \oplus W_{2\ldots N})\sig (W_1^T \oplus W_{2\ldots
N}^T) + (S_{smo} W_1 \oplus  W_{2\ldots N})\sig (W_1^T S_{smo}^T
\oplus W_{2\ldots N}^T)] \nonumber \\ &=& \det[(W_1 \oplus
 W_{2\ldots N})\sig (W_1^T \oplus W_{2\ldots N}^T) +
(W_1 W_1^{-1} S_{smo} W_1 \oplus  W_{2\ldots N}) \sig (W_1^T
S_{smo}^T W_1^{-1 T}W_1^T \oplus  W_{2\ldots N}^T)] \nonumber \\ &=&
\det\{[W_1 \oplus W_{2\ldots N}][\sig + (W_1^{-1} S_{smo} W_1
\id_{2\ldots N}) \sig (W_1^{-1} S_{smo} W_1\oplus \id_{2\ldots
N})^T][W_1 \oplus  W_{2\ldots N}]^T\} \nonumber \\ &=&  \det[\sig +
(W_1^{-1} S_{smo} W_1 \oplus  \id_{2\ldots N}) \sig (W_1^{-1}
S_{smo} W_1\oplus \id_{2\ldots N})^T] \; ,
\end{eqnarray}
\end{widetext}
where we exploited the group properties of
$Sp(2,\R)$, the fact that a symplectic operation $S$
has $\det S=1$, and the property that the inverse $S^{-1}$
of a symplectic transformation $S$
is itself symplectic. Now, from the cyclic property of the trace,
it follows that
$W_1^{-1} S_{smo} W_1$ is itself a traceless symplectic operation,
i.e.~a single-mode operation of the form \eq{m}. Thus the minimum of
the above determinant, taken over the
entire set {$S_{smo}(\alpha,\beta)$} of single-mode unitary operations,
is invariant under local symplectic operations $W_1
\oplus W_{2\ldots N}$ performed on state $\sig$. Thus, without loss
of generality, we can choose a pure $N$-mode Gaussian state in
the phase-space Schmidt form $\sig_W$ of \eq{schm}. Therefore,
\begin{eqnarray}
\label{split}
&& \det[\sig_W+(S_{smo}\oplus \id_{2\ldots N})\sig_W
(S_{smo}\oplus \id_{2\ldots N})^T] \nonumber \\
&& = 2^{2(N-1)} [(a^2-1)^2+4 a^2 \beta^2] \, .
\end{eqnarray}
The minimum is then acquired, as $\beta \ge \sqrt{\alpha^2+1}$,
for $\beta=1,\,\alpha=0$. The corresponding extremal single-mode
operation is then, finally
\begin{equation}
\label{smop}
S_{smo}(0,1) \equiv \Sigma_2 \equiv \omega=\left(
                                        \begin{array}{cc}
                                          0 & 1 \\
                                          -1 & 0 \\
                                        \end{array}
                                      \right)\; .
\end{equation}
This is a simple rotation of $\pi/2$ in phase space, and may be seen
as the CV analogue of the spin-flip operation on qubits,
realized by the $\sigma_y$ Pauli matrix.

We observe that a product state, characterized by a CM in direct sum
form, $\sig^\oplus \equiv \sig_1 \oplus \sig_{2\ldots N}$, is left
{\em invariant} by the extremal single-mode operation:
\begin{equation}\label{tensjens}[S_{smo}(0,1) \oplus \id_{2\ldots N}] \sig^\oplus [S_{smo}(0,1)
\oplus \id_{2\ldots N}]^T = \sig^\oplus.\end{equation} That is, on
pure product Gaussian states, extremal and invariant (or preserving)
operations coincide, in full analogy with the finite-dimensional
case analyzed in Ref. \cite{squo}. Hence, {\em a pure Gaussian
states is separable if and only if there exists a traceless
single-mode symplectic (unitary) operation that leaves it
unperturbed.} This is again in perfect analogy with the
discrete-variable analysis performed for qubits and qutrits
\cite{squo}.
\begin{figure}[b]
\centering{\includegraphics[width=8.5cm]{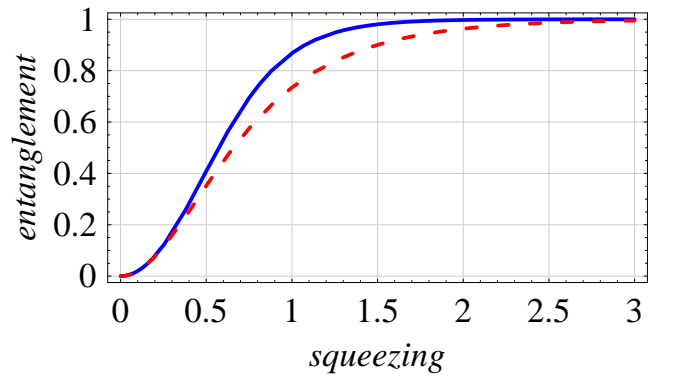}%
\caption{\label{ficompa} (color online) Entanglement of pure ($1
\times (N-1)$)-mode Gaussian states as a function of the single-mode
squeezing $r$, defined such that $a=\cosh{2r}$. The solid line
depicts the distance-based geometric measure $D$ defined in
\eq{dopt}, while the dashed line corresponds to the linear entropy
of entanglement $E_L$, \eq{ella}.}}
\end{figure}
The minimum distance \eq{distanza}, achieved for
$\beta=1,\,\alpha=0$, can now be evaluated explicitly and
reads
\begin{equation}
\label{dopt}
D(\sig)=1-\frac{2^N}{2^{N-1}(a^2+1)}=\frac{a^2-1}{a^2+1} \; .
\end{equation}
The quantity $D(\sig)$ is {\em a measure} of the entanglement
between mode $1$ and the rest of the system, being an increasing
function of the single-mode mixedness factor $a \ge 1$. For
product states $a=1$ and one correctly retrieve $D(\sig^\oplus)=0$.
One should recall that \eq{dopt} holds
in general, and not only for states in Schmidt form, once $a$ is
identified with the (locally invariant) symplectic eigenvalue of the
reduced CM $\sig_1$ of mode $1$, computable from its determinant as
$a=\sqrt{\det\sig_1}$. From this observation, it follows immediately
that $D(\sig)$, even if constructed through the action of
local unitaries, is invariant under them -- as already proved by
\eq{longproof} -- and is thus a proper entanglement measure.

The $1 \times (N-1)$ linear entropy of entanglement for the state $\sig$
(corresponding to the tangle for qubits \cite{ckw}) reads
\begin{equation}
\label{ella}
E_L(\sig) \, = \, \frac{a-1}{a} \; .
\end{equation}
We see that $E_L$ is a monotonic function of $D$, thus qualitatively equivalent
to but yet not exactly coincident with the latter everywhere, at variance with
the discrete-variable case, in which they do strictly coincide \cite{squo}
(The behavior of the two entanglement monotones is compared in Fig.~\ref{ficompa}).
The fact that the linear entropy of the reduced state does not coincide
exactly with the minimum distance achieved under local symplectic
operations may be traced back to the non uniqueness in the definition
of the ``tangle'' for Gaussian states of CV systems.
For qubits, at least four different definitions coalesce
into the same entanglement monotone: (i) squared concurrence \cite{ckw};
(ii) local linear entropy \cite{osborneverstraete}; (iii) squared
negativity (negativity equals concurrence for pure qubit states
\cite{frankconc}); (iv) minimum distance under single-qubit unitary
transformation \cite{squo}. On the other hand, while the concurrence
is not well defined in CV systems, the other definitions of the
tangle all give rise to different (yet equivalent) entanglement
quantifiers in these systems. For instance, the Gaussian tangle
defined as the squared negativity \cite{hiroshima}, in analogy with
definition (iii), reads
\begin{equation}
\label{gtau}
\tau_G(\sig)= [a^2 + a \sqrt{a^2 - 1}  - 1]/2 \; .
\end{equation}
The von Neumann entropy of entanglement, for reference, is given by
$E_V(\sig)= [(a+1)/2] \log [(a+1)/2] - [(a-1)/2] \log [(a-1)/2]$.
All these measures are monotonically increasing functions of each
other (and of $a$), some of them being normalized between $0$ and
$1$ (like $D$ and $E_L$), the others diverging in the
limit of infinite squeezing, $a \rightarrow \infty$.

\section{Experimental remarks and future perspectives}

The minimum distance $D$ provides a new {\em bona fide} geometric
measure of entanglement for pure Gaussian states, close in spirit to
the low-dimensional, discrete-variable counterpart introduced in
Ref.~\cite{squo}, and embodying yet another generalization of the
tangle. However, we would like to remark that, among the three
possible CV versions of the tangle, only $\tau_G$, \eq{gtau},
satisfies the CV generalization of the Coffman-Kundu-Wootters
monogamy inequality \cite{ckw,osborneverstraete}, as proved in
Ref.~\cite{hiroshima} for all, pure and mixed, $N$-mode Gaussian
states.

On the other hand, the geometric measure of entanglement $D(\sig)$
that we have introduced in this work for pure Gaussian states, has
the nice property of being amenable to direct experimental
investigation, once  two copies of an unknown Gaussian state with CM
$\sig$ are available. One first needs a phase plate in order to
rotate one copy of $\pi/2$, realizing the operation $\Sigma_2$, as
demonstrated e.g.~in \cite{francamentemeneinfischio}. Thereafter,
the evaluation of the overlap between the rotated copy and the
unrotated one involves standard tools of linear optics, as routinely
demonstrated in the determination of the fidelity ${\cal F}$ of
teleportation experiments with continuous variables
\cite{furutelep}, or in the implementation of interferometric
schemes \cite{oi} that can be realized even without homodyning
\cite{fiurasek04}. Our result thus suggests a way to the direct
measurement of CV entanglement in pure $1\times N$ Gaussian states,
in analogy with what achieved experimentally in the case of qubits:
In that case, the entanglement, quantified by the two-point
concurrence, has been directly measured on the two-fold copy of
unknown two-qubit pure states \cite{walborn}.

This proposal looks  especially appealing for Gaussian states with a
small number of modes. A relevant example is provided by  three-mode
Gaussian states, whose CM assumes in general the following
expression in terms of $2$ by $2$ submatrices,
\begin{equation}\sig = \left(\begin{array}{ccc}
\sig_{1} & \gr\varepsilon_{12} & \gr\varepsilon_{13} \\
\gr\varepsilon_{12}^{\sf T} & \sig_{2} & \gr\varepsilon_{23} \\
\gr\varepsilon_{13}^{\sf T} & \gr\varepsilon_{23}^{\sf T} & \sig_{3} \\
\end{array}\right) \; . \label{subma} \end{equation}
The structural and informational properties of three-mode Gaussian
states, with a special emphasis on the pure-state instance, have
been extensively studied in Ref.~\cite{3modi}, while a scheme for
their production via interlinked nonlinear interactions in
$\chi^{(2)}$ media has been presented in Ref.~\cite{ferraro},
together with preliminary experimental results. When modes $2$ and
$3$ have the same average number $\bar n$ of thermal photons, the
corresponding (parametric) pure three-mode Gaussian state is said to
be `bisymmetric' and its CM can be written in the standard form of
\eq{subma}, with
\begin{eqnarray*}
  \sig_1 &=& a \id_2,\\
  \sig_2 = \sig_3 &=&  \left(\frac{a+1}{2}\right) \id_2, \\
  \gr\varepsilon_{23} &=& \left(\frac{a-1}{2}\right) \id_2, \\
  \gr\varepsilon_{12} = \gr\varepsilon_{13} &=& {\rm
  diag}\left\{\frac{\sqrt{a^2-1}}{\sqrt{2}},\,-\frac{\sqrt{a^2-1}}{\sqrt{2}}\right\},
  \\ \\
  \mbox{and}\quad a &=& 4 \bar n + 1.
  \label{basseps}
\end{eqnarray*}
The geometric entanglement between the first mode and the group of
modes $2$ and $3$, as obtainable from the single-mode perturbation
\eq{smop} applied on mode $1$, is then directly given by \eq{dopt}
as a function of $\bar n$. The three-mode Gaussian states of this
family are known to be optimal resources for $1 \rightarrow 2$  CV
telecloning (i.e.~cloning at distance, or equivalently teleportation
to more than one receiver) of single-mode coherent states
\cite{telecloning}, as discussed also in \cite{ferraro, 3modi}. The
single-clone fidelity $\cal F$ exhibits a non-monotonic, concave
behaviour as a function of $\bar n$, reaching the maximum ${\cal
F}^{\max}=2/3$ at $\bar n=1/2$. Very recently, the first
experimental demonstration of unconditional $1 \rightarrow 2$
telecloning of unknown coherent states has been realized
\cite{exptelecloning}, with a measured fidelity for each clone of
${\cal F} = 0.58 \pm 0.01$ (surpassing the classical threshold of
$0.5$). This experimental milestone has raised renewed interest
towards CV quantum communication \cite{press}. In the context of
this work, such an achievement entails that all the elementary steps
required to access pure-state Gaussian entanglement from a geometric
point of view have been already successfully undertaken. Our
prescription, therefore, is likely to be seen ``at work''
experimentally on multimode Gaussian states in the near future.

In this paper we have dealt with {\em pure} Gaussian states only. It
is natural to ask whether a suitable generalization of the
present approach is able to provide a geometric interpretation,
possibly amenable to direct experimental tests, of {\em mixed}-state
entanglement measures as well. In this
respect, it is important to clarify to which extent the results of
this paper are still valid for mixed states. {\em In primis}, it is
generally true that the extremal single-mode operation, \eq{smop},
preserves tensor product, even mixed Gaussian states [see
\eq{tensjens}]. However, convex combination of product states,
i.e.~separable mixed Gaussian states, are not left invariant by the
action of such local operation. Accordingly, the corresponding
geometric measure (minimum distance $D$) defined by \eq{dopt} is
not, in general, an entanglement monotone for mixed Gaussian states.
One can thus conclude that, in the mixed-state scenario, the mere
action of $S_{smo}(0,1)$ leads to a distinction between tensor
product states (totally uncorrelated, on which the distance
\eq{dopt} is zero) and states displaying some form of (quantum
and/or classical) correlation. A refinement is henceforth necessary
in order to aim at discriminating, from a geometric point of view,
the quantum portion -- entanglement -- from the total amount of
correlations. A feasible way to deal with this issue seems that of
identifying a minimal set of single-mode unitary operations, such
that a suitably defined ``distance'' involving their combined
action, may turn to be equivalent (or to provide bounds) to known
entanglement monotones (e.g.~negativities, tangles and/or measures
based on the Gaussian convex roof \cite{ordering}). One should then
be able to readily provide a recipe for the practical estimation of
mixed-state Gaussian entanglement with few local measurements (see
also \cite{Rigolin}).

Finally, we would like to remark that the framework introduced
in Ref. \cite{squo}, and further discussed in the present paper,
can be naturally applied to investigate criticality and entropy
scaling in the ground states of harmonic lattices \cite{arealaw},
with the purpose of establishing connections similar to those unveiled
for the ground states of spin systems at criticality \cite{faberverruca}.
This interesting perspective will be the object of further future studies.

\acknowledgments

We acknowledge financial support from MIUR under PRIN National
Project 2005, INFN, CNR-INFM Coherentia, CNISM-CNR, and ISI
Foundation. G. A. is grateful to M. Aspachs, A. Boada and G. Bracons
for their kind hospitality while this paper was in production.

\end{document}